\documentstyle[twoside,fleqn,espcrc2,epsf]{article}

\newcommand{\AmS}{{\protect\the\textfont2
  A\kern-.1667em\lower.5ex\hbox{M}\kern-.125emS}}

\title{P-wave heavy-light mesons using NRQCD and D234}

\author{Randy Lewis\address{Department of Physics,
        University of Regina, Regina, SK, Canada S4S 0A2}
        and 
        R. M. Woloshyn\address{TRIUMF, 4004 Wesbrook Mall, Vancouver, BC,
        Canada V6T 2A3}}
       
\begin{document}

\begin{abstract}
The masses of S- and P-wave heavy-light mesons are computed in quenched QCD
using a classically and tadpole-improved action on anisotropic lattices.
Of particular interest are the splittings among P-wave states, which have not
yet
been resolved experimentally; even the ordering of these states continues to
be discussed in the literature.  The present work leads to upper bounds
for these splittings, and is suggestive, but not conclusive, about the
ordering.
\end{abstract}

\maketitle

\section{INTRODUCTION}

The masses of S-wave heavy-light mesons have been a valuable testing ground
for lattice NRQCD studies.  The calculations are precise enough to test
convergence of the NRQCD expansion\cite{LewWol}, and the effects of
quenching\cite{Coletal}.  Also, experimental measurements are precise enough
to constrain the values of coefficients in the NRQCD action\cite{Coletal}.

For P-waves, the situation is quite different.  Experiments have not yet
resolved the splittings among P-wave states, and Isgur\cite{Isg} has recently
promoted the suggestion, initiated long ago by Schnitzer\cite{Sch}, that
the P-wave states may be inverted relative to the conventional hydrogen-like
ordering.
This inversion is predicted by using the experimental $K^*$ masses and mixing 
angle as input,
then extrapolating from the constituent strange quark to charm or bottom via a
$1/M$-expanded nonrelativistic potential.
Verification of this prediction would thus lend support to the treatment
of a constituent strange quark as ``heavy'' within a nonrelativistic quark 
model.

To date, quenched lattice QCD calculations have not seen an 
inversion\cite{MicPei,Boy,Alietal}, although the uncertainties are often
large enough to make firm conclusions difficult.

The present work is an exploration of this physics using an improved action
on anisotropic lattices.  The results obtained below agree qualitatively
with previous lattice calculations, but a quantitative comparison reveals
differences.

\section{ACTION}

The lattice action has three terms: gauge action, light quark action
and heavy quark action.  
The entire action is classically and tadpole-improved with
the tadpole factors, $U_{0,s}$ and $U_{0,t}$, defined as the mean links in 
Landau gauge in a spatial and temporal direction, respectively.

The gauge action includes a sum over 1$\times$2 rectangular
plaquettes as well as 1$\times$1 elementary plaquettes, and therefore the
leading classical errors are quartic in lattice spacing.
For light quarks, a D234 action\cite{AKL} is used with parameters set
to their tadpole-improved classical values.  Its leading errors are
cubic in lattice spacing.  

The heavy quark action is NRQCD\cite{NRQCD}, which is discretized to give 
the following Green's function propagation\cite{LewWol}:
\begin{eqnarray}
G_{\tau+1} \!\!\!&=&\!\!\! \left(1-\frac{a_tH}{2n}\right)^n
    \frac{U_4^\dagger}{U_{0,t}}
    \left(1-\frac{a_tH}{2n}\right)^n G_\tau,
\end{eqnarray}
with $n=5$ chosen for this work, and the
Hamiltonian is truncated at $O(1/M)$,
\begin{eqnarray}
H &=& \frac{-\Delta^{(2)}}{2M}
                 -\frac{1}{U_{0,s}^4}\frac{g}{2M}\mbox{{\boldmath$\sigma$}}
                    \cdot\tilde{\bf B} + \frac{{a_s}^2\Delta^{(4)}}{24M}, \\
\tilde{B}_i &=& \frac{1}{2}\epsilon_{ijk}\tilde{F}_{jk}.
\end{eqnarray}
A tilde denotes removal of the leading discretization errors, leaving the
action with quadratic lattice spacing errors.
Notice that the coefficients in $H$ are set to their
tadpole-improved classical values.

\section{RESULTS}

All data presented here come from 2000 gauge field configurations on
10$^3\times$30 lattices at
$\beta=2.1$ with a bare aspect ratio of $a_s/a_t=2$, which corresponds to
$a_t = 0.10$ fm.
The light quark mass is fixed by $\kappa=0.24$, which gives
$m_\pi/m_\rho = 0.52\pm0.01$, ie. $m_q \sim m_s/2$.
Fixed time boundaries are used for the light quark propagators so they fit
naturally into a meson with an NRQCD heavy quark propagator.

Four heavy quark masses have been studied: $a_sM = 1, 3, 5, \infty$.
Calculation of the kinetic mass of the ${}^1S_0$ heavy-light meson
leads to the following charm and bottom quark masses: $a_sM_c \sim 1.2$
and $a_sM_b \sim 5$.  More precise determinations of these parameters
are not required for the present exploratory study.

Familiar heavy-light meson creation operators are used,
\begin{equation}
   \sum_{\vec x}Q^\dagger(\vec{x})\Omega(\vec{x})\Gamma(\vec{x})q(\vec{x}),
\end{equation}
where $\Omega(\vec{x})$ is given in table~\ref{operators} and the smearing
operator is
\begin{equation}
   \Gamma(\vec{x}) = [1+c_s\Delta^{(2)}(\vec{x})]^{n_s}.
\end{equation}
All plots shown here use $(c_s,n_s)=(0.15,10)$ at the source
(which is fixed to timestep 4) and a local sink.
\begin{table}
\caption{Heavy-light meson creation operators.}\label{operators}
\begin{tabular}{rl}
${}^{2S+1}L_J$ & $\Omega(\vec{x})$ \\
\hline
${}^1S_0$ & (0,$I$) \\
${}^3S_1$ & (0,$\sigma_i$) \\
${}^1P_1$ & (0,$\Delta_i$) \\
${}^3P_0$ & (0,$\sum_i\Delta_i\sigma_i$) \\
${}^3P_1$ & (0,$\Delta_i\sigma_j-\Delta_j\sigma_i$) \\
${}^3P_2$ & (0,$\Delta_i\sigma_i-\Delta_j\sigma_j$) or \\
          & (0,$\Delta_i\sigma_j+\Delta_j\sigma_i$), $i \neq j$
\end{tabular}
\vspace*{-5mm}
\end{table}

\begin{figure}[hb]
\epsfxsize=180pt \epsfbox[30 470 498 700]{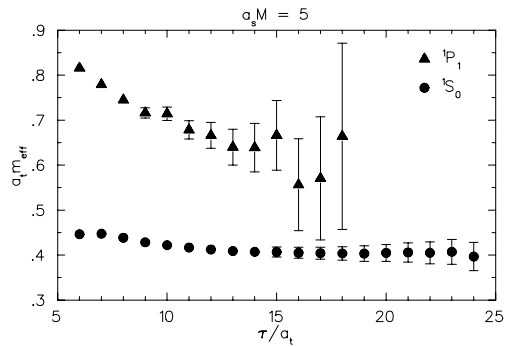}
\caption{The ${}^1S_0$ and ${}^1P_1$ simulation energies in lattice units at 
$a_sM=5$.}\label{sp}
\end{figure}
Fig. \ref{sp} shows the ${}^1S_0$ and ${}^1P_1$ simulation energies
in the bottom region.  In NRQCD, only energy differences are physical but
this plot is an indication of the data quality.
As for all plots in this work, the uncertainties are determined from 5000
bootstrap ensembles.
It should be noted that the ${}^1P_1$-${}^1S_0$ splitting is in agreement
with previous lattice determinations (eg. \cite{MicPei,Boy,Alietal}).

\begin{figure}[hb]
\epsfxsize=180pt \epsfbox[30 470 498 680]{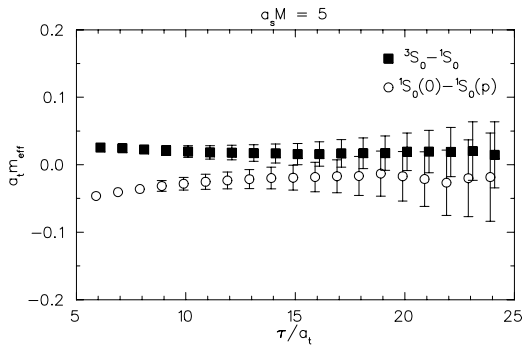}
\caption{The S-wave mass splittings in lattice units at 
$a_sM=5$.  The nonzero momentum ``$p$'' is $2\pi/(10a_s)$.}\label{sdiff}
\end{figure}
Fig. \ref{sdiff} shows the ${}^3S_1$-${}^1S_0$ mass splitting and the
kinetic shift of the ${}^1S_0$ simulation energy in the bottom region.
In physical units, the ${}^3S_1$-${}^1S_0$ splitting is found to be 
34$\pm$5~MeV, in agreement with the known quenched result 
(eg. \cite{LewWol,MicPei,Boy,Alietal}).

\begin{figure}[htb]
\epsfxsize=180pt \epsfbox[30 370 498 720]{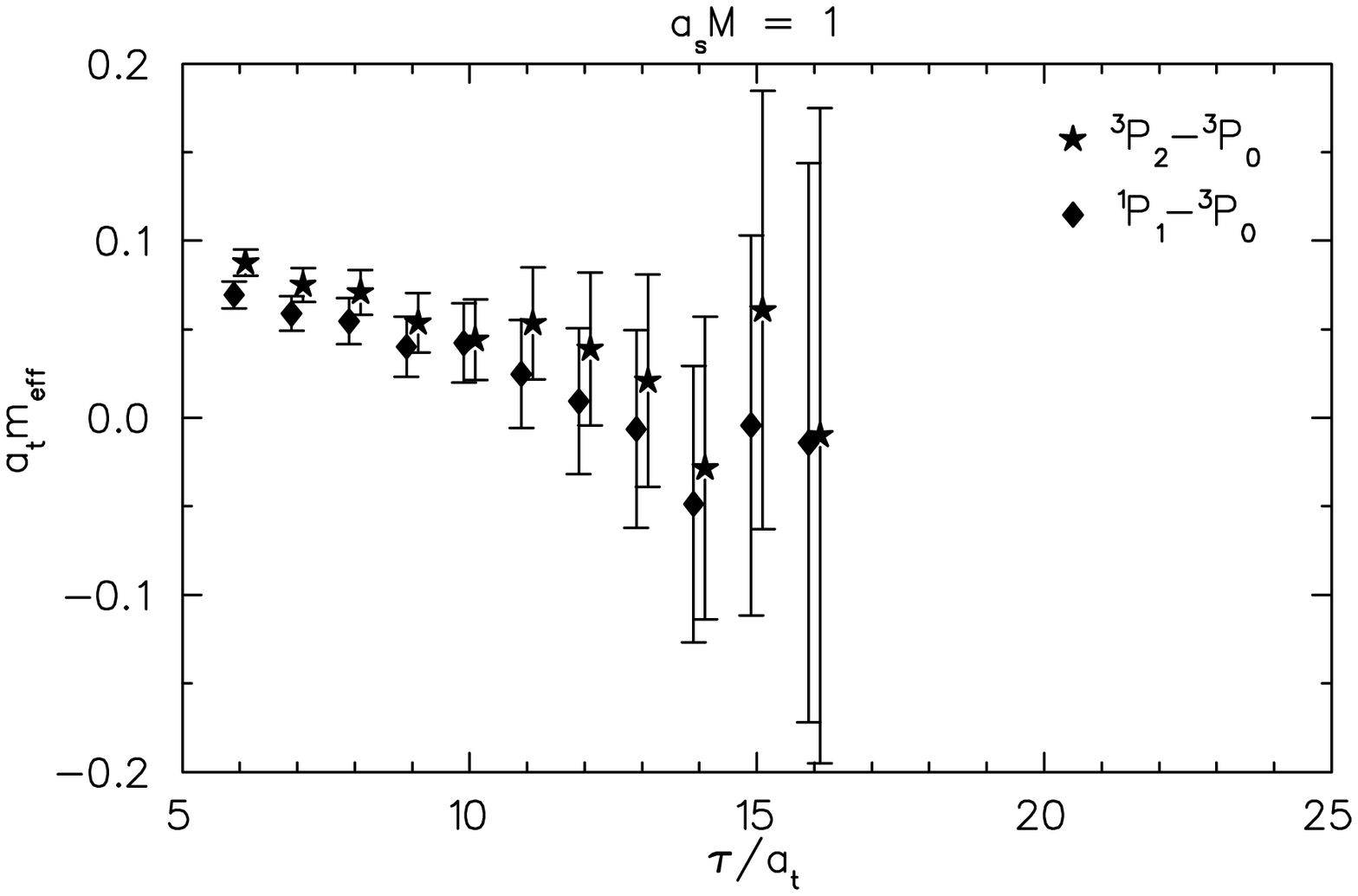}
\epsfxsize=180pt \epsfbox[30 370 498 720]{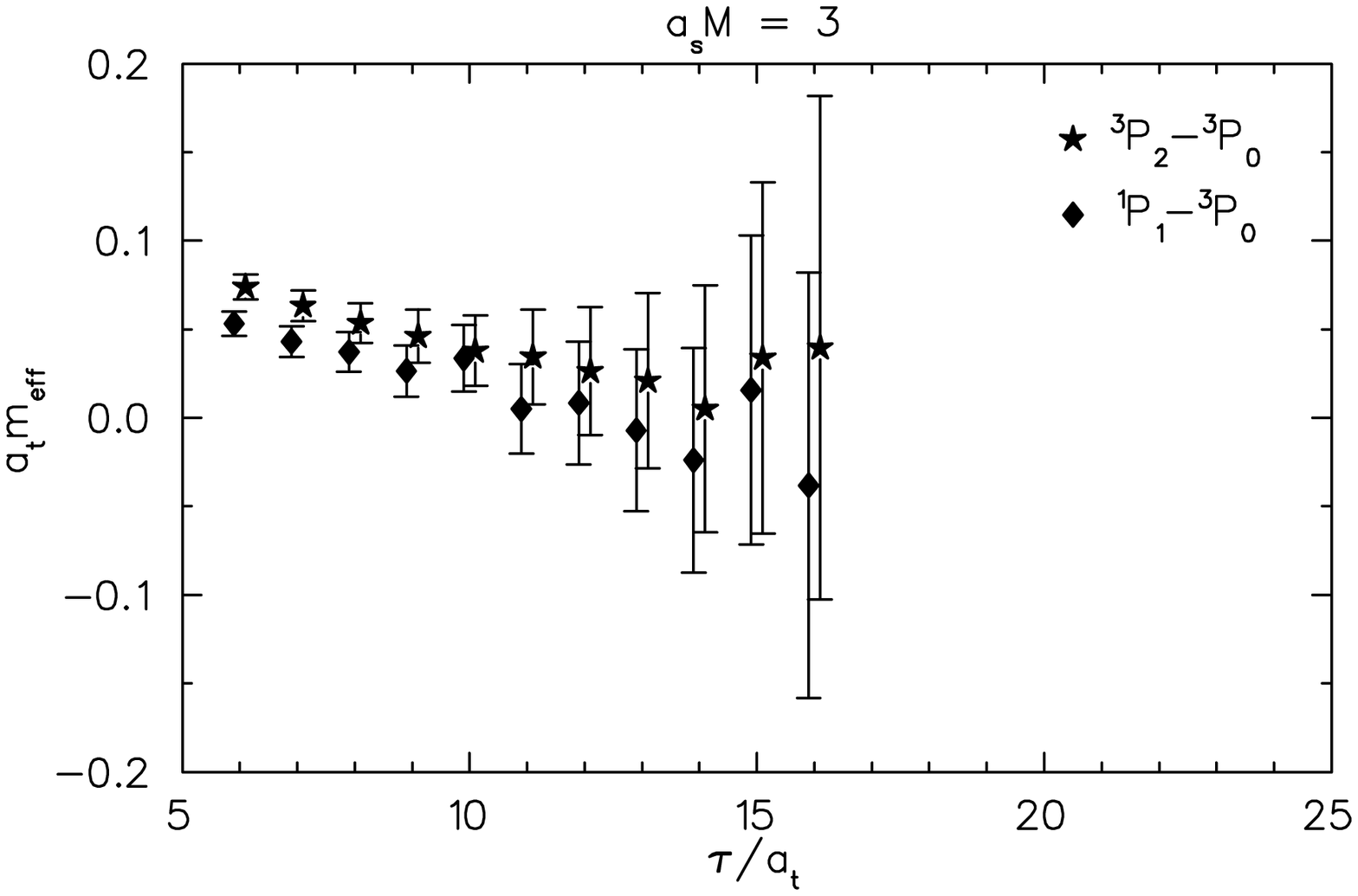}
\epsfxsize=180pt \epsfbox[30 370 498 720]{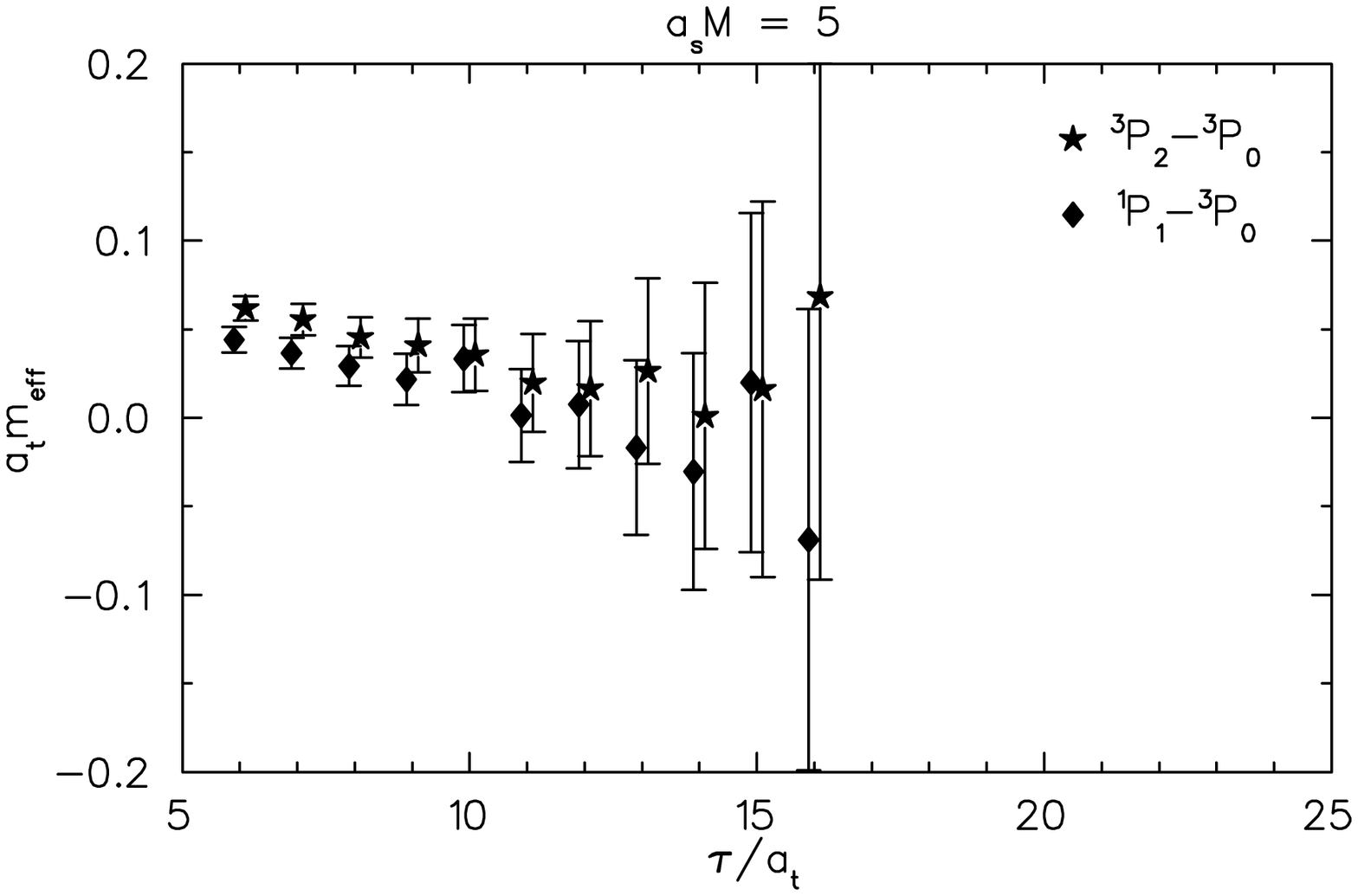}
\epsfxsize=180pt \epsfbox[30 470 498 720]{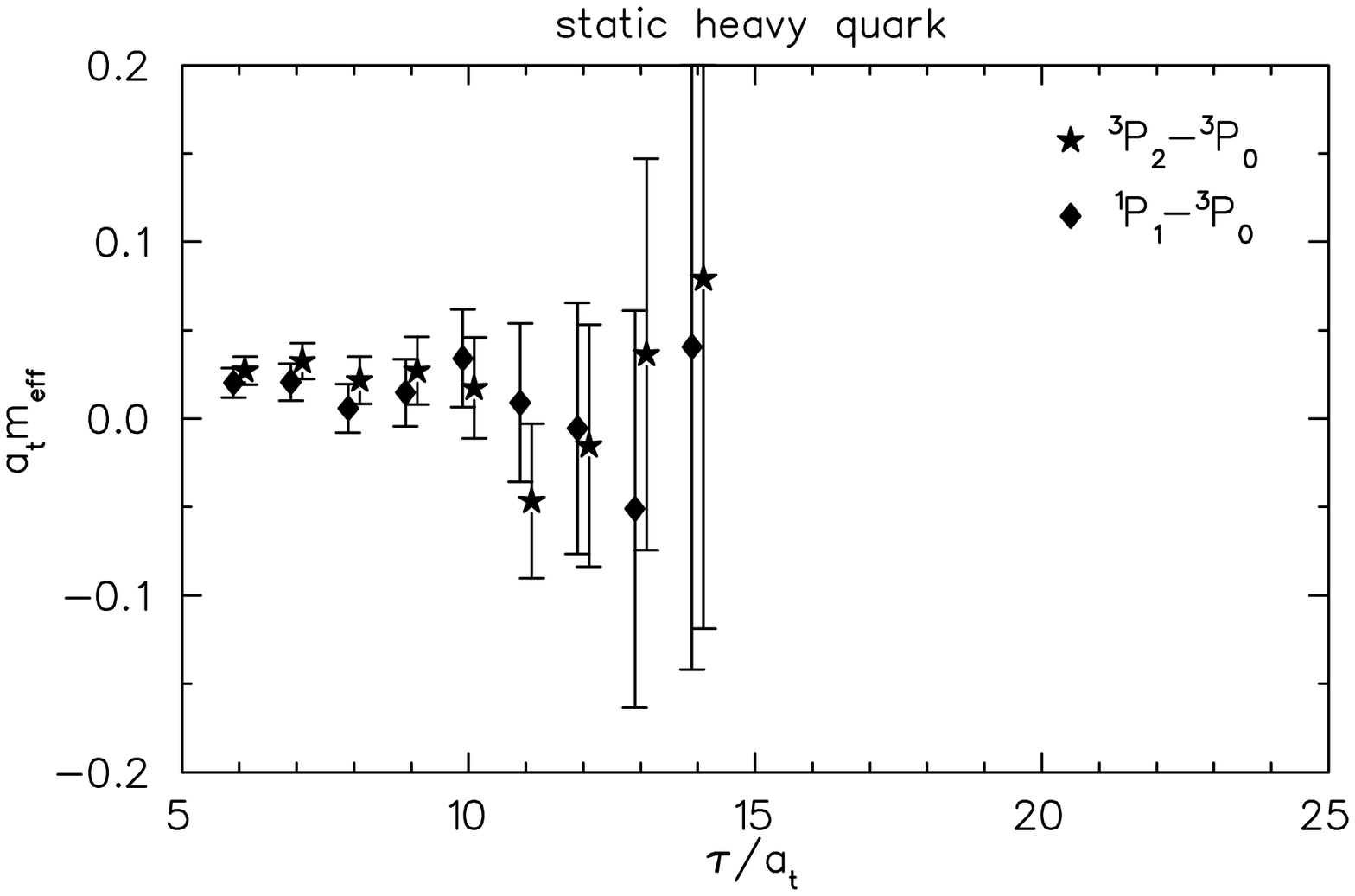}
\caption{The P-wave mass splittings in lattice units.}\label{pdiff}
\end{figure}
Fig. \ref{pdiff} displays the mass splittings among P-wave states at
all available quark masses.  The ${}^3P_1$ is not shown since it
is not distinguishable from the ${}^1P_1$.  
Near the source, the states are organized according to the familiar
hydrogen-like pattern: $P_0$, $P_1$, $P_2$ from lightest to heaviest.  

As time increases, the uncertainties grow such that one might
be tempted to define a plateau which begins rather close to the source.
However, the central values of the mass splittings tend to decrease
for increasing time, so a negative mass splitting (i.e. an inverted
spectrum) cannot be ruled out conclusively.

Using $a_t = 0.10$ fm, Fig. \ref{pdiff} indicates that the 
${}^3P_2$-${}^3P_0$ splitting for bottom mesons
is less than 100~MeV, which should be contrasted
with Ref.~\cite{Alietal}, where lattice NRQCD gave 183$\pm$34~MeV.
While it is true that the two determinations stem from different methods
(different lattice actions, anisotropic versus isotropic lattices,
different light quark masses, slightly different temporal lattice spacings,
\ldots), it is disconcerting that the final results are not in better
agreement.

In conclusion, this work has led to an upper bound for the ${}^3P_2$-${}^3P_0$ 
(bottom) splitting which lies below a previous lattice determination.
An inverted spectrum, although not suggested by the calculation, 
cannot yet be ruled out.

\section*{ACKNOWLEDGEMENTS}

R.L. thanks A. Ali Khan, P. Boyle, S. Collins and J. Hein for discussions.
This work was supported in part by the Natural Sciences and Engineering
Research Council of Canada.

\end{document}